\documentclass{article}
\usepackage{spconf,amsmath,graphicx}
\usepackage{cite}
\usepackage{amsthm,amsmath,amssymb,bm}
\usepackage{amsfonts}
\usepackage{booktabs}
\usepackage{url}
\usepackage{xcolor}
\usepackage{multirow}
\usepackage{makecell}
\usepackage[caption=false]{subfig}
\usepackage{float}
\usepackage[colorlinks,linkcolor=red]{hyperref}

\usepackage{bbding}
\newcommand{\vModelName}{\textit{AudioSR}}

\title{AudioSR: Versatile Audio Super-resolution at Scale}

\name{Haohe Liu$^1$, Ke Chen$^2$, Qiao Tian$^3$, Wenwu Wang$^1$, Mark D. Plumbley$^1$}
\address{
  $^1$Centre for Vision Speech and Signal Processing, University of Surrey, \\
  $^2$University of California San Diego,  \\
  $^3$Speech, Audio \& Music Intelligence (SAMI), ByteDance \\
}

\begin{document}
\maketitle

\begin{abstract}

Audio super-resolution is a fundamental task that predicts high-frequency components for low-resolution audio, enhancing audio quality in digital applications. Previous methods have limitations such as the limited scope of audio types~(e.g., music, speech) and specific bandwidth settings they can handle~(e.g., $4$ kHz to $8$ kHz). In this paper, we introduce a diffusion-based generative model, \vModelName, that is capable of performing robust audio super-resolution on versatile audio types, including sound effects, music, and speech. Specifically, \vModelName~can upsample any input audio signal within the bandwidth range of $2$~kHz to $16$~kHz to a high-resolution audio signal at $24$~kHz bandwidth with a sampling rate of $48$~kHz. Extensive objective evaluation on various audio super-resolution benchmarks demonstrates the strong result achieved by the proposed model. In addition, our subjective evaluation shows that \vModelName~can acts as a plug-and-play module to enhance the generation quality of a wide range of audio generative models, including AudioLDM, Fastspeech2, and MusicGen. Our code and demo are available at \url{https://audioldm.github.io/audiosr}. 


\end{abstract}


\section{Introduction}


Audio super-resolution~(SR) aims to estimate the higher-frequency information of a low-resolution audio signal, which yields a high-resolution audio signal with an expanded frequency range. High-resolution audio signals usually offer a better listening experience, which is often referred to as high fidelity. Due to the ability to enhance audio signal quality, audio super-resolution plays a significant role in various applications, such as historical recording restoration~\cite{liu2021voicefixer}.

Previous studies on audio SR have primarily focused on specific domains, with a particular emphasis on speech SR. Early research decompose the speech SR task into spectral envelope estimation and excitation generation \cite{kontio2007neural}. Recent works employing deep learning techniques, such as AECNN \cite{heming-towards-sr-wang2021towards}, NuWave \cite{lee2021nu}, and NVSR \cite{liu2022neural}, have shown superior performance compared to traditional methods. In addition to speech, there have been efforts to address music SR, including studies on general music \cite{hu2020phase} and specific instruments \cite{rakotonirina2021self}.

Apart from the limited scope of audio, existing research on audio SR also has primarily been conducted in controlled experimental settings, limiting its applicability in real-world scenarios. 
An important challenge in audio super-resolution, as highlighted in \cite{liu2022neural}, is the issue of bandwidth mismatch. This occurs when the bandwidth of the test data differs from that of the training data, leading to model failure. However, this issue has not received significant attention in the literature, as previous works typically assume consistent bandwidth settings for both training and testing data. 
In practice, the input bandwidth of test audio can vary due to factors such as limitations in recording devices, sound characteristics, or applied compression processes. 
Only a few studies have explored flexible input bandwidth, including NVSR \cite{liu2022neural} and NuWave2 \cite{han2022nu}. However, these methods still primarily focus on speech SR without generalizing to a broader domain.

\begin{figure}[t]
    \centering
    \vspace{-2mm}
    \includegraphics[width=0.9\linewidth]{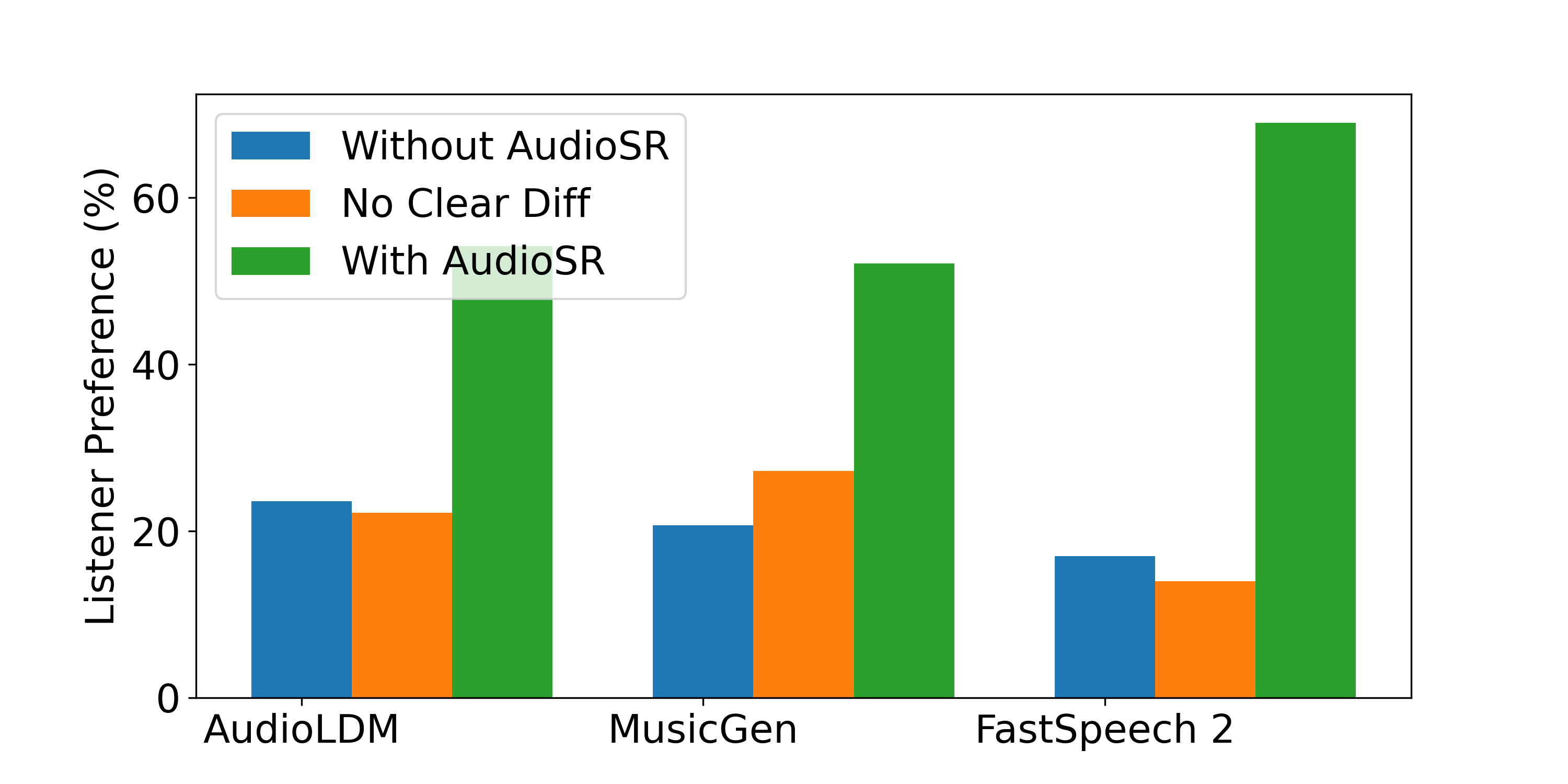}
    \vspace{-5mm}
    \caption{Subjective evaluation shows that applying~\vModelName~for audio super-resolution on the output of audio generation models can significantly enhance the perceptual quality.}
    \label{fig:preference}
    \vspace{-5mm}
\end{figure}

In this paper, we propose a novel method that addresses the limitations of previous work on limited audio types and controlled sampling rate settings. We introduce a method called \vModelName, which extends audio SR to a general domain, including all audible sounds such as music, speech, and sound effects. Moreover, \vModelName~is capable of handling a flexible input sampling rate between $4$kHz and $32$kHz, covering most of the use cases in real-world scenarios. It has been found that the prior knowledge learned by the neural vocoder is helpful for reconstructing higher frequency components in audio SR tasks~\cite{liu2022neural}. Therefore, \vModelName~follows \cite{liu2022neural} to perform audio SR on the mel-spectrogram and utilizes a neural vocoder to synthesize the audio signal. To estimate the high-resolution mel-spectrogram, we follow AudioLDM~\cite{liu2023audioldm} to train a latent diffusion model on learning the conditional generation of high-resolution mel-spectrogram from low-resolution mel-spectrogram. Our experiment demonstrates that~\vModelName~has achieved promising SR results on speech, music, and sound effects with different input sampling rate settings. Our subjective evaluation on enhancing the output of text-to-audio model AudioLDM~\cite{liu2023audioldm}, text-to-music model MusicGen~\cite{copet2023simple-musicgen}, and text-to-speech model Fastspeech2~\cite{Fastspeech2} show that~\vModelName~can be a plug-and-play module for most audio generation models to enhance listening quality. Our contributions are summarized as follows:

\begin{list}{\labelitemi}{\leftmargin=1em \itemsep=0pt}
\item Our proposed \vModelName~is the first system to achieve audio SR in the general audible audio domain, covering various types of audio such as music, speech, and sound effects. 
\item \vModelName~can handle a flexible audio bandwidth ranging from $2$kHz to $16$kHz, and extend it to $24$kHz bandwith with $48$kHz sampling rate.
\item Besides the promising results on audio SR benchmarks, \vModelName~is also verified to be a plug-and-play module for enhancing the audio quality of various audio generation models such as AudioLDM, MusicGen, and FastSpeech2.
\end{list}

The paper is organized as follows. Section~\ref{sec:problem-formulation} provides a general formulation of the audio super resolution task. Section~\ref{sec:methods} provides a detailed explanation of the design of~\vModelName. 
The detailed experimental setup is discussed in Section~\ref{sec:experiments}. Our experimental results are presented in Section~\ref{sec:result}, and we conclude the paper in Section~\ref{sec:conclusion-final}.

\section{Problem Formulation}
\label{sec:problem-formulation}

Given an analog signal that has been discretely sampled at a rate of $l$ samples per second, resulting in a low-resolution sequence of values $x_l = [x_i]_{i=1,2,...T\cdot l}$, the goal of audio super-resolution (SR) is to estimate a higher resolution signal $y_h = [y_i]_{i=1,2,...T\cdot h}$ sampled at a rate of $h$ samples per second, where $h>l$ and $T$ is the total duration in seconds. According to Nyquist's theory, $x_l$ and $y_h$ have maximum frequency bandwidths of $l/2$ Hz and $h/2$ Hz respectively. Therefore, the information contained between frequencies of $h/2-l/2$ Hz is missing from $x_l$, and estimating this ``missing'' frequency data is the core objective of the SR task.

In this paper, we follow the method proposed in NVSR~\cite{liu2022neural} to decompose the original audio SR task into two steps, including \textit{(i) High-resolution Mel spectrogram Estimation}, and \textit{(ii) Mel Spectrogram to Waveform Reconstruction with a Neural Vocoder.} 
Specifically, we first resampling $x_l$ to $x_h$ using cubic interpolation, where $x_h$ has a higher sampling rate $h$ but with limited maximum bandwidth of $l/2$~Hz. we follow the steps in~\cite{liu2022neural} to calculate the mel spectrogram of both $x_h$ and $y_h$, resulting $X_{m\times n}$ and $Y_{m\times n}$, respectively, where $m$ is the number of time frames and $n$ is the number of mel frequency bins. Then we utilize a generative model to learning the process of estimating $Y$ based on $X$, which is denoted as $\mathcal{G}_{\theta}:X\mapsto \hat{Y}$, where $\theta$ are the parameters of model $\mathcal{G}$. Finally, a neural vocoder is employed to reconstruct the high sampling rate audio signal based on the estimation of $Y$, which can be formulated as $\mathcal{V}_{\phi}:\hat{Y}\mapsto\hat{y}_h$, where $\mathcal{V}$ is the neural vocoder and $\phi$ are the learnable parameters. 

\section{Method}
\label{sec:methods}

\begin{figure}[tbp]
    \centering
    \includegraphics[width=1.0\linewidth]{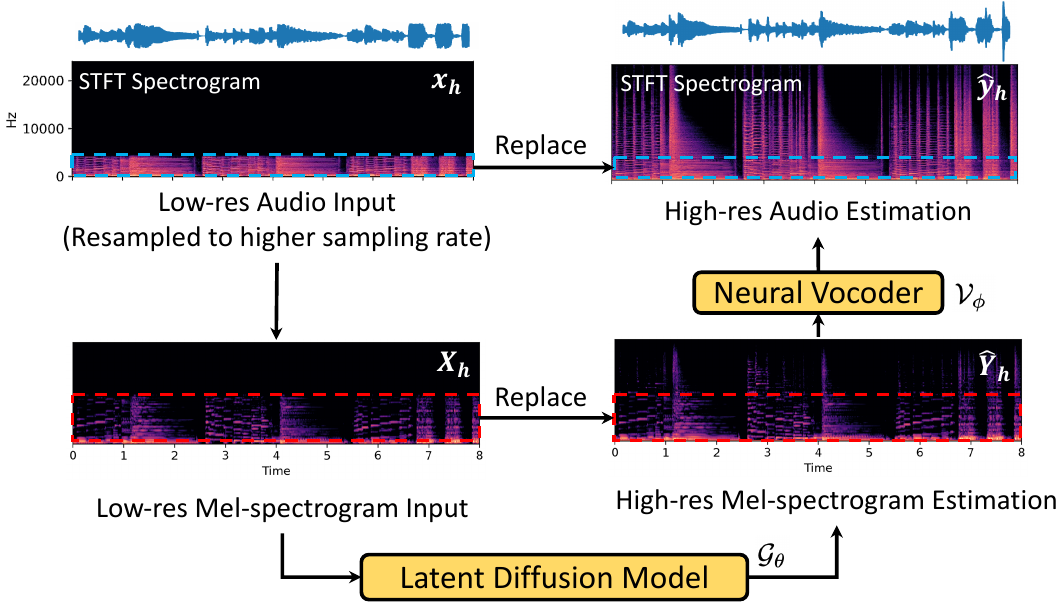}
    \caption{The~\vModelName~architecture. The replacement-based post-processing aims to preserve the original lower-frequency information in the model output. }
    \label{fig:main}
    \vspace{-5mm}
\end{figure}

The architecture of the proposed \vModelName~is demonstrated in Figure~\ref{fig:main}. After resampling the low-resolution audio $x_l$ to $x_h$, the system first calculates both the STFT spectrogram and the mel spectrogram of $x_h$. Note that the higher frequency bins in $X_h$ are empty because $x_h$ does not have high-frequency information. $X_h$ is then used as a conditioning signal to guide the pre-trained latent diffusion model to estimate the high-resolution mel spectrogram $\hat{Y}_h$. To ensure consistency in the low-frequency information between $X_h$ and $\hat{Y}_h$, we replace the lower frequency part of $\hat{Y}_h$ with that of $X_h$. 
The mel-spectrogram after low-frequency replacement serves as the input to the neural vocoder, which output applies a similar technique to replace the low-frequency information with that of the input low-resolution audio. 
We introduce the training of the latent diffusion model and neural vocoder in Section~\ref{sec:latent-diffusion-model}. The post-processing algorithm is elaborated in Section~\ref{sec:post-processing}.

\subsection{High-resolution Waveform Estimation}
\label{sec:latent-diffusion-model}

\textbf{Latent diffusion model~(LDM)} has demonstrated promising results in various domains, including image synthesis~\cite{rombach2022high-stablediffusion} and audio generation~\cite{liu2023audioldm}. In this study, we employ the LDM to estimate high-resolution mel-spectrograms. The training of our LDM is conducted within a latent space learned by a pre-trained variational autoencoder~(VAE) $\mathcal{F(\cdot)}$. The VAE is trained to perform autoencoding with a small compressed latent space in the middle, denoted as $\mathcal{F}:X \mapsto z_0 \mapsto \hat{X}$. By leveraging the lower-dimensional representation $z_0$, the LDM can learn the generation of $z_0$ instead of $X$, resulting in a substantial reduction in computational cost. We adopt the methodology proposed in AudioLDM to optimize the VAE model, including the use of reconstruction loss, Kullback–Leibler divergence loss, and discriminative loss.

We follow the formulation introduced in AudioLDM~\cite{liu2023audioldm} to implement the LDM, with improvements on the training objective, noise schedule, and conditioning mechanism. It has been found that the common noise schedule used in the diffusion model is flawed~\cite{lin2023common}, particularly because the noise schedule in the final diffusion step of LDM does not correspond to a Gaussian distribution. To address this issue, we follow~\cite{lin2023common} to update the noise schedule to a cosine schedule. This adjustment ensures that a standard Gaussian distribution can be achieved at the final diffusion step during training. Additionally, we incorporate the velocity prediction objective~\cite{salimans2022progressive} on reflection of using the new noise schedule. The final training objective of our LDM is 
\begin{equation}
    \text{argmin}_{\mathcal{G}_{\theta}} || v_{k} - \mathcal{G}(z_k, k, \mathcal{F}_{\text{enc}}(X_{l}); \theta) ||^{2}_2,
    \label{eq:diffusion-objective}
\end{equation}
where $z_k$ represents the data of $z_0$ at diffusion step $k\in[1,...,K]$, $||\cdot||_2$ denotes the Euclidean distance, $\mathcal{F}_{\text{enc}}$ denote the VAE encoder, and as described in~\cite{lin2023common}, $v_k$ is calculated based on $z_0$, representing the prediction target of $\mathcal{G}$ at time step $k$. We adopt the Transformer-UNet architecture proposed in~\cite{liu2023audioldm2} as $\mathcal{G}$. The input to $\mathcal{G}$ is obtained by concatenating $z_k$ with the $\mathcal{F}_{\text{enc}}(X_{l})$, which is the VAE latent extracted from the low-resolution mel-spectrogram $X_{l}$. To incorporate classifier-free guidance, following the formulation in~\cite{liu2023audioldm}, we replace $\mathcal{F}_{\text{enc}}(X_{l})$ with an empty tensor at a random rate (e.g., $10\%$) during training. After training the latent diffusion model, we perform sampling using the DDIM sampler~\cite{song2020denoising-ddim}.

\noindent
\textbf{Neural Vocoder.}~The LDM is capable of estimating high-resolution mel spectrograms. However, since mel-spectrograms are not directly audible, we employ a neural vocoder based on HiFiGAN~\cite{kong2020hifi} to convert the mel-spectrograms into waveforms. To address the issue of spectral leakage when implementing the original HiFiGAN, we adopt the multi-resolution discriminator~\cite{you2021gan} into the HiFiGAN vocoder. We optimize the vocoder using diverse audio data, as discussed in Section~\ref{sec:datasets}, resulting in a vocoder that operates at a sampling rate of $48$kHz and can work on diverse types of audio.

\subsection{Post-processing and Pre-processing}
\label{sec:post-processing}

\textbf{Post-processing.}~The input low-resolution audio features $X_h$ and $x_h$ are identical to the lower frequency bands in the estimation target, $Y_h$ and $y_h$. As a result, we can reuse the available information from $X_h$ and $x_h$ to enhance both the LDM output $\hat{Y}_h$ and neural vocoder output $\hat{y}_h$. To accomplish this, we first determine the $0.99$ roll-off frequency $c$ of the entire input audio based on an open-source method\footnote{\url{https://librosa.org/doc/main/generated/librosa.feature.spectral_rolloff.html}} applied to both $X_h$ and the STFT spectrogram of $y_h$. Subsequently, we replace the spectrogram components below the cutoff frequency in the LDM output $\hat{Y}_h$ and vocoder output $\hat{y}_h$, with the corresponding information in the $X_h$ and $x_h$, respectively. This post-processing method can ensure the final output does not significantly alter the lower-frequency information.

\noindent
\textbf{Pre-processing.}~
To minimize the mismatch between model training and evaluation, we perform preprocessing to the input audio during evaluation with a lowpass-filtering operation. We use the same method in post-processing to calculate the $0.99$ roll-off frequency and perform lowpass filtering with an order $8$ \textit{Chebyshev} filter.

\section{Experiment}
\label{sec:experiments}

\begin{table*}[t!]
\vspace{-8mm}
\centering
\scriptsize
\begin{tabular}{ccccccccccc|cc}
\hline
\multicolumn{11}{c|}{Objective Evaluation}                                                                                                                                                                                                  & \multicolumn{2}{c}{Subjective Evaluation}                    \\ \hline
\multicolumn{4}{c|}{VCTK (Speech)}                                                    & \multicolumn{4}{c|}{\multirow{2}{*}{AudioStock (Music)}}                              & \multicolumn{3}{c|}{\multirow{2}{*}{ESC-50 (Sound Effect)}} & \multicolumn{2}{c}{\multirow{2}{*}{ESC-50~($4$kHz Cutoff Freq)}} \\
Cutoff-frequency & $4$kHz          & $8$kHz          & \multicolumn{1}{c|}{$12$kHz}         & \multicolumn{4}{c|}{}                                                                 & \multicolumn{3}{c|}{}                                       & \multicolumn{2}{c}{}                                         \\ \cline{1-4}
GT-Mel           & $0.64$ & $0.64$ & \multicolumn{1}{c|}{$0.64$}                                          & Cutoff-frequency & $4$kHz          & $8$kHz          & \multicolumn{1}{c|}{$16$kHz}         & $4$kHz               & $8$kHz               & $16$kHz             & System                       & Overall Quality               \\ \cline{5-13} 
Unprocessed      & $5.15$          & $4.85$          & \multicolumn{1}{c|}{$3.84$}          & GT-Mel           & $0.61$ & $0.61$ & \multicolumn{1}{c|}{$0.61$}                                          & $0.84$ & $0.84$ & \multicolumn{1}{c|}{$0.84$}                                   & GT-Mel                       & $4.35$                          \\
NuWave~\cite{lee2021nu}           & $1.42$          & $1.36$          & \multicolumn{1}{c|}{$1.22$}          & Unprocessed      & $4.25$          & $3.48$          & \multicolumn{1}{c|}{$1.99$}          & $3.90$                & $3.07$               & $2.25$              & Unprocessed                  & $3.01$                          \\
NVSR~\cite{liu2022neural}             & $\mathbf{0.91}$ & $\mathbf{0.81}$ & \multicolumn{1}{c|}{$0.70$}           & NVSR-DNN         & $1.67$          & $1.49$          & \multicolumn{1}{c|}{$1.13$}          & $\mathbf{1.64}$      & $1.59$               & $1.76$              & NVSR-DNN                     & $2.84$                          \\
\vModelName          & $1.30$           & $1.11$          & \multicolumn{1}{c|}{$0.94$}          & NVSR-ResUNet     & $1.70$           & $1.34$          & \multicolumn{1}{c|}{$0.95$}          & $1.80$                & $1.69$               & $1.67$              & NVSR-ResUNet                 & $3.16$                          \\
\vModelName\textit{-Speech}   & $1.03$          & $0.82$          & \multicolumn{1}{c|}{$\mathbf{0.69}$} & \vModelName          & $\mathbf{0.99}$ & $\mathbf{0.74}$ & \multicolumn{1}{c|}{$\mathbf{0.73}$} & $1.74$               & $\mathbf{1.57}$      & $\mathbf{1.35}$     & \vModelName                      & $\mathbf{4.01}$                 \\ \hline
\end{tabular}
\caption{Objective and subjective evaluation results for 48kHz audio SR of speech, music, and sound effect data with varying cutoff frequencies in the input audio. The objective metric used for evaluation is the LSD, where lower values indicate superior performance. The subjective metric measures the overall listening quality, with higher values indicating better performance.}
\label{tab:main}
\end{table*}

\begin{figure*}
    \centering
    \includegraphics[width=1.0\linewidth]{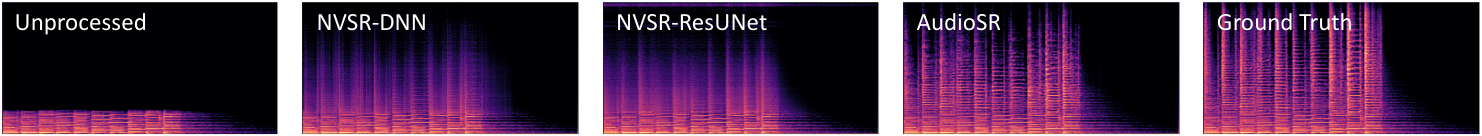}
    \caption{Comparison of different systems. \vModelName~performs significantly better than the baseline NVSR models.}
    \label{fig:example}
    \vspace{-5mm}
\end{figure*}

\label{sec:datasets}

\noindent
\textbf{Training Datasets.} The datasets used in this paper include MUSDB18-HQ~\cite{musdb18-hq}, MoisesDB~\cite{pereira2023moisesdb}, MedleyDB~\cite{bittner2014medleydb}, FreeSound~\cite{mei2023wavcaps}\footnote{\url{https://labs.freesound.org/}}, and the speech dataset from OpenSLR\footnote{\url{https://openslr.org/}}, which are downloaded by following the link provided by VoiceFixer~\cite{liu2021voicefixer}. All the audio data used are resampled at $48$kHz sampling rate. The total duration of the training data is approximately $7000$ hours. We utilize all these datasets to optimize VAE, LDM, and HiFi-GAN.

\noindent
\textbf{Training Data Simulation.} We follow the method introduced in NVSR~\cite{liu2022neural} to simulate low-high resolution audio data pairs. Given a high-resolution audio data $y_h$, we first perform lowpass filtering to the audio with a cutoff frequency uniformly sampled between $2$kHz and $16$kHz. To address the filter generalization problem~\cite{heming-towards-sr-wang2021towards}, the type of the lowpass filter is randomly sampled within \textit{Chebyshev, Elliptic, Butterworth, and Boxcar}, and the order of the lowpass filter is randomly selected between $2$ and $10$. 

\noindent
\textbf{Evaluation Datasets.} We performed both subjective and objective evaluations. For subjective evaluations, we adopt the output of MusicGen~(caption from MusicCaps~\cite{agostinelli2023musiclm}), AudioLDM~(caption from AudioCaps~\cite{kim2019audiocaps}), and Fastspeech2~(transcription from LJSpeech~\cite{ljspeech17}) to study if the~\vModelName~can enhance the quality of the generation. For MusicGen we use audio tagging~\footnote{\url{https://github.com/kkoutini/PaSST}} to filter out the non-musical generation output. Finally, we collected $50$ samples from MusicGen, $50$ samples from AudioLDM, and $20$ samples from FastSpeech2, and processed them with~\vModelName~for subjective evaluations on listener preference. Besides, we curate three benchmarks for objective evaluation, including ESC50~(sound effect)~\cite{piczak2015esc}, AudioStock~(music)\footnote{\url{https://audiostock.net/}}, and VCTK~(speech)~\cite{liu2022neural}.  The AudioStock dataset is built by hand-picking $100$ high-quality music with $10$ different genres. We only use the fold-5 in the ESC50 dataset as the evaluation set. 

\noindent
\textbf{Evaluation Metrics} For objective evaluation, we adopt the LSD metric, as used in prior studies~\cite{heming-towards-sr-wang2021towards,liu2022neural}. Following the setup of~\cite{liu2023audioldm2}, we conduct two types of subjective evaluation on Amazon Mturk\footnote{\url{https://www.mturk.com/}}: Overall quality rating and preference comparison. In the overall quality rating, raters assign a score between $1$ and $5$ to reflect the audio quality. In the preference comparison, raters compare two audio files and select the one that sounds better. 
\label{sec:evaluation-metrics}

\vspace{-3mm}
\section{Result}
\label{sec:result}

We trained two versions of \vModelName~for evaluation: the basic \vModelName~that works on arbitrary audio types and input sampling rates, and a speech data fine-tuned variant called \vModelName\textit{-Speech}. Our primary baseline for comparison is NVSR~\cite{liu2022neural}, which employs a similar mel-spectrogram and vocoder-based pipeline for audio SR tasks. The main distinction between \vModelName~and NVSR lies in the mel-spectrogram estimation approach: \vModelName~utilizes a latent diffusion model, while NVSR employs either a multilayer perceptron (NVSR-DNN) or a residual UNet (NVSR-ResUNet). For speech SR, we also compare with NuWave~\cite{lee2021nu} as a baseline model, which also employs a diffusion model for audio SR.

As shown in Table~\ref{tab:main},~\vModelName~has achieved promising results on both objective and subjective evaluation. For music SR,~\vModelName~achieves state-of-the-art performance across all cutoff frequency settings, outperforming the baseline NVSR model by a large margin. For speech SR, \vModelName\textit{-Speech} achieves the best performance on the $24$kHz to $48$kHz upsampling task. Also, the comparison between~\vModelName~and~\vModelName\textit{-Speech} indicates that finetuning on a small domain of data can significantly improve the LSD. 

The LSD metric does not always align with perceptual quality. In the $8$kHz (i.e., $4$kHz cutoff frequency) to 48kHz upsampling task on the ESC-50 dataset, we observed that NVSR-DNN achieved the best performance with an LSD score of $1.64$. However, subjective evaluations indicated that the perceptual quality of NVSR-DNN the worst with a score of $2.84$, significantly lower than \vModelName's score of $4.01$. These findings suggest that LSD may not be a suitable evaluation metric for audio SR tasks on sound effect data, warranting further investigation in future research.

As depicted in Figure~\ref{fig:preference}, our subjective preference test demonstrates that the utilization of~\vModelName~significantly enhances the perceptual quality of the AudioLDM, MusicGen, and FastSpeech2 output. It is worth noting that the output of MusicGen is already in a high sampling rate of $32$kHz, which may contribute to the relatively high rate of ``No Clear Difference'' responses. However,~MusicGen still exhibits a significantly improved perceptual quality after applying ~\vModelName. 

\vspace{-3mm}
\section{Conclusion}
\label{sec:conclusion-final}
This paper presents \vModelName, a $48$kHz audio super-resolution model that is capable of working with diverse audio types and arbitrary sampling rate settings. Through evaluation of multiple audio super-resolution benchmarks, \vModelName~demonstrates superior and robust performance on various types of audio and sampling rates. Additionally, our subjective evaluation highlights the effectiveness of \vModelName~in enabling plug-and-play quality improvement for the audio generation models, including AudioLDM, MusicGen, and Fastspeech2. Future work includes extending \vModelName~for real-time applications and exploring appropriate evaluation protocols for audio super-resolution in the general audio domain.

\vspace{-3mm}
\section{Acknowledgments}
This research was partly supported by the British Broadcasting Corporation Research and Development, Engineering and Physical Sciences Research Council (EPSRC) Grant EP/T019751/1 ``AI for Sound'', and a PhD scholarship from the Centre for Vision, Speech and Signal Processing (CVSSP), Faculty of Engineering and Physical Science (FEPS), University of Surrey. For the purpose of open access, the authors have applied a Creative Commons Attribution (CC BY) license to any Author Accepted Manuscript version arising. 

\label{sec:conclusion}

\bibliographystyle{IEEEtran}
\bibliography{strings}

\end{document}